\documentclass[12pt]{article}
\usepackage{graphicx}
\usepackage{color, rotating, amsmath, amsfonts , subfig , color}
\usepackage[justification=centering]{caption}

\usepackage{lineno,hyperref}
\usepackage{setspace}

\newcommand{\vsp}{\vspace*{3mm}}

\newcommand{\be}{\begin{equation}}
\newcommand{\ee}{\end{equation}}
\newcommand{\bd}{\begin{displaymath}}
\newcommand{\ed}{\end{displaymath}}
\newcommand{\bea}{\begin{eqnarray}}
\newcommand{\eea}{\end{eqnarray}}
\newcommand{\bean}{\begin{eqnarray*}}
\newcommand{\eean}{\end{eqnarray*}}

\newcommand{\one}{1\!\!{\rm I}}

\newcommand{\order}{{\mathcal O}}

\newcommand{\bnull}{\mbox{\boldmath $0$}}

\newcommand{\R}{{\rm I\!R}}

\newcommand{\sig}{\mathbf{\Sigma}}

\newcommand{\bx}{\mbox{\protect\boldmath $x$}}

\newcommand{\bA}{\mbox{\protect\boldmath $A$}}

\newcommand{\bC}{\mbox{\protect\boldmath $C$}}

\newcommand{\bM}{\mbox{\protect\boldmath $M$}}

\newcommand{\bQ}{\mbox{\protect\boldmath $Q$}}

\newcommand{\bS}{\mbox{\protect\boldmath $S$}}

\newcommand{\bX}{\mbox{\protect\boldmath $X$}}

\newcommand{\btheta}{\mbox{\protect\boldmath $\theta$}}

\newcommand{\bmu}{\mbox{\protect\boldmath $\mu$}}

\newcommand{\bLambda}{\mbox{\protect\boldmath $\Lambda$}}

\newcommand{\bXi}{\mbox{\protect\boldmath $\Xi$}}
\newcommand{\rmd}{{\rm d}}
\newcommand{\rme}{{\rm e}}

\newcommand{\pr}{\boldsymbol{\Lambda_z}}

\newcommand{\pra}{\mathbf{A}}

\newcommand{\I}{\mathbb{I}_d}

\newcommand{\dhalf}{\frac{d}{2}}

\newcommand{\pop}{\mathbf{\Sigma}}

\usepackage[mathscr]{euscript}

\newcommand{\Data}{\mathscr{D}}

\begin{document}

\begin{center}{\Large\bf Accurate Bayesian Data Classification\\[1mm] without Hyperparameter Cross-validation}\end{center}
\vspace*{2mm}

\begin{center}{\large M Sheikh and ACC Coolen\\[2mm]
{\small King's College London, U.K.}}
\end{center}
\vsp

\begin{abstract}
\noindent
We extend the standard Bayesian multivariate Gaussian generative data classifier by considering a generalization of the conjugate, normal-Wishart prior distribution and by deriving the hyperparameters analytically via evidence maximization. The behaviour of the optimal hyperparameters is explored in the high-dimensional data regime. The classification accuracy of the resulting generalized model is competitive with state-of-the art Bayesian discriminant analysis methods, but without the usual computational burden of cross-validation.\\
\end{abstract}

\noindent
\begin{center}{\small
}\end{center}


\section{Introduction}

In the conventional formulation of classification problems one aims to map data samples $\bx\in\R^d$ correctly into  discrete classes $y\in\{1,\ldots C\}$, by inferring the underlying statistical regularities from a given training set $\Data=\{(\bx_1,y_1),\ldots,(\bx_n,y_n)\}$ of i.i.d. samples and corresponding  classes. The standard approach to this task is to define a suitable parametrization $p(\bx,y|\btheta)$ of the multivariate distribution from which the samples in $\Data$ were drawn. If the number of samples $n$ is large compared to the data dimensionality $d$, computing point estimates of the unknown parameters $\btheta$ by maximum likelihood (ML) or maximum a posteriori probability (MAP) methods is accurate and usually sufficient. On the other hand, if the ratio $d/n$ is not small, point estimation based methods are prone to overfitting. This is  the `curse of dimensionality'.
Unfortunately, the regime of finite $d/n$ is quite relevant for medical applications, where clinical data-sets often report on relatively few patients but contain many measurements per patient\footnote{This is the case for rare diseases, or when obtaining tissue material is nontrivial or expensive, but measuring extensive numbers of features in such material (e.g. gene expression data) is relatively simple and cheap.}.

In generative classification models, a crucial role is played by the class-specific sample covariance matrices, that capture the correlations between the components of $\bx$. These will have to be inferred, either explicitly or implicitly.
While the sample covariance matrix $\pop$  is a consistent estimator for the population covariance matrix $\pop_0$ in the limit $d/n\to 0$, for finite $d/n$  the empirical covariance eigenvalue distribution $\varrho(\xi)$ is a poor estimator for its population counterpart $\varrho_0(\lambda)$. 
This becomes more pronounced as $d/n$ increases\footnote{While $\varrho(\lambda)$ is not a good estimator for $\varrho_0(\lambda)$, \cite{jonsson1982some} showed that in contrast $\int\!\rmd\lambda~\varrho(\lambda)\lambda$ is a good estimate of
$\int\!\rmd\lambda ~\varrho_0(\lambda)\lambda$; the bulk spectrum becomes more biased as $d/n$ increases, but the sample eigenvalue  {\em average} does not.}. 
In additional to the clear bias in covariance estimation induced by high ratios of $d/n$, the geometry of high-dimensional spaces produces further extreme and often counterintuitive values of probability masses and densities \cite{mackay1999comparison}.

The overfitting problem has been known for many decades, and 
many strategies have been proposed to combat its impact on multivariate point estimation inferences.
Regularization methods add a penalty term to the ML loss function. The penalty strength acts as a hyperparameter, and is to be estimated. The penalty terms punish e.g. increasing values of  the sum of absolute parameter values (LASSO) or squared parameter values (ridge regression), or a linear combination of these (elastic net)  \cite{zou2005regularization}. They appear naturally upon introducing prior probabilities in Bayesian inference, followed by MAP point estimation. 
Feature selection methods seek to identify a subset of `informative' components of the sample vectors. They range from linear methods such as Principal Component Analysis \cite{hotelling1933analysis} to non-linear approaches such as auto-encoders \cite{hinton2006reducing}.  They can guide experimental work, by suggesting which data features to examine. Most of these techniques use heuristics to determine the number of features to select.
Early work by \cite{stein1956inadmissibility} introduced the concept of shrinking a traditional estimator toward a `grand average'. In the univariate case, this was an average of averages \cite{efron1977stein}. For the multivariate case, the James-Stein estimator is an admissible estimator of the population covariance matrix \cite{james1961estimation}. This idea was further developed by \cite{ledoit2004well,haff1980empirical}.
More recent approaches use mathematical tools from theoretical physics to predict (and correct for) the overfitting bias in ML regression analytically \cite{Coolen_etal_2017}.

Any sensible generative model  for classifying vectors in $\R^d$ will have at least $\order(d)$ parameters. 
The fundamental cause of overfitting is the fact that in high-dimensional spaces, where $d/n$ is finite even if $n$ is large, the posterior parameter distribution $p(\btheta|\Data)$ (in a Bayesian sense) will be extremely sparse. Replacing this posterior  by a delta-peak, which is what point estimation implies,  is always a very poor approximation, irrespective of which protocol is used for estimating the location of this peak. It follows that by   avoiding point estimation altogether, i.e. by retaining the full posterior distribution  and doing all integrations over model parameters {\em analytically}, one should reduce overfitting effects, potentially allowing for high-dimensional data-sets to be classified reliably. Moreover, only hyperparameters will then have to be estimated (whose dimensionality is normally small, and independent of $d$), so one avoids the prohibitive computational demands of sampling high-dimensional spaces. 
The need to do all parameter integrals analytically  limits us in practice to parametric generative models with class-specific multivariate Gaussian distributions. Here the model parameters to be integrated over are the class means in $\R^d$  and class-specific $d\times d$ covariance matrices, and with carefully chosen priors one can indeed obtain analytical results. The Wishart distribution is the canonical prior for the covariance matrices. Analytically tractable choices 
for the class means are the conjugate \cite{keehn1965note,geisser1964posterior} or the non-informative priors  \cite{brown1999discrimination,srivastava2006distribution}. 

As the data dimensionality increases, so does the role of Bayesian priors and their associated hyperparameters, and the method used for computing hyperparameters impacts more on the performance of otherwise identical models. The most commonly used route for hyperparameter estimation appears to be cross-validation. This requires re-training one's model $k$ times for $k$-fold cross-validation; for leave-one-out cross-validation, the model will need to be re-trained $n$ times. 

In this paper we  generalize the family of prior distributions for  parametric generative models with class-specific multivariate Gaussian distributions, without loss of analytical tractability, and we compute hyperparameters via evidence maximization, rather than cross-validation. This allows us to derive closed form expressions for the predictive probabilities of two special model instances. The numerical complexity of our approach does not increase significantly with $d$ since all integrals whose dimensions scale with $d$ are solved analytically.  

In section \ref{sec:model} we first define our generative Bayesian classifier and derive the relevant integrals. Special analytically solvable cases of these integrals, leading to two models (A and B), are described in section \ref{sec:hyper} along with the evidence maximization estimation of hyperparameters. Closed form expressions for the predictive probabilities corresponding to these two models are obtained in section \ref{sec:predictive}. We then examine the behaviour of the hyperparameters in section \ref{sec:behave}, and carry out comparative classification performance tests on synthetic and real data-sets in section \ref{sec:results}. We conclude our paper with a discussion of the main results. 

\section{Definitions}
\label{sec:model}
\subsection{Model and objectives}

We have data $\Data=\{(\bx_1,y_1),\ldots,(\bx_n,y_n)\}$ consisting of $n$ samples of pairs $(\bx,y)$, where $\bx\in\R^d$ is a vector of covariates, and $y\in\{1,\ldots,C\}$ a discrete outcome label.
We seek to predict the outcome $y_0$  associated with a {\em new} covariate vector $\bx_0$, given the data $\Data$. 
So we want to compute
\begin{eqnarray}
p(y_0|\bx_0,\Data)&=& \frac{p(y_0,\bx_0|\Data)} {\sum_{y=1}^C p(y,\bx_0|\Data)}
~=~ \frac{p(y_0,\bx_0|\bx_1,\ldots,\bx_n;y_1,\ldots,y_n)} {\sum_{y=1}^C p(y,\bx_0|\bx_1,\ldots,\bx_n;y_1,\ldots,y_n)}
\nonumber
\\
&=& \frac{p(\bx_0,\ldots,\bx_n;y_0,\ldots,y_n)} {\sum_{y=1}^C p(\bx_0,\ldots,\bx_n;y,y_1,\ldots,y_n)}
\end{eqnarray}
We next need an expression for the joint distribution $p(\bx_0,\ldots,\bx_n;y_0,\ldots,y_n)$.
We assume that all pairs $(\bx_i,y_i)$ are drawn independently from a parametrized distribution $p(\bx,y|\btheta)$ whose parameters $\btheta$ we don't know. Using de Finetti's representation theorem and the fact that exchangeability is a weaker condition than i.i.d, we can write the joint distribution of $\{ (\bx_i,y_i) \}_{i=0}^n$ as 
\begin{eqnarray}
p(\bx_0,\ldots,\bx_n;y_0,\ldots,y_n)&=& \int\!\rmd\btheta ~p(\btheta)\prod_{i=0}^n p(\bx_i,y_i|\btheta)
\end{eqnarray}
It now follows that 
\begin{eqnarray}
p(y_0|\bx_0,\Data)&=& \frac{ \int\!\rmd\btheta ~p(\btheta)\prod_{i=0}^n p(\bx_i,y_i|\btheta)} {\sum_{y=1}^C  \int\!\rmd\btheta ~p(\btheta)p(\bx_0,y|\btheta)\prod_{i=1}^n p(\bx_i,y_i|\btheta)}
\end{eqnarray}
We regard all model parameters with dimensionality that scales with the covariate dimension $d$ as {\em micro-parameters}, over which we need to integrate (in the sense of $\btheta$ above). Parameters with $d$-independent dimensionality are regarded as {\em hyperparameters}.  The hyperparameter values  will be called a `model' $H$. Our equations will now acquire a label $H$:
\begin{eqnarray}
p(y_0|\bx_0,\Data,H)&=& \frac{ \int\!\rmd\btheta ~p(\btheta|H)\prod_{i=0}^n p(\bx_i,y_i|\btheta,H)} {\sum_{y=1}^C  \int\!\rmd\btheta ~p(\btheta|H)p(\bx_0,y|\btheta,H)\prod_{i=1}^n p(\bx_i,y_i|\btheta,H)}
\label{eq:inference_formula}
\end{eqnarray}
The Bayes-optimal hyperparameters $H$ are those that maximise the evidence, i.e. 
\begin{eqnarray}
\hat{H}&=& {\rm argmax}_H p(H|\Data)~=~{\rm argmax}_H \log\Big\{\frac{p(\Data|H)p(H)}{\sum_{H^\prime}p(\Data|H^\prime)p(H^\prime)}\Big\}
\nonumber
\\&=&{\rm argmax}_H \Big\{ \log \int\!\rmd\btheta ~p(\btheta|H)\prod_{i=1}^n p(\bx_i,y_i|\btheta,H)+\log p(H)\Big\}
\label{eq:maximise_evidence}
\end{eqnarray}
What is left is to specify the parametrization $p(\bx,y|\btheta)$ of the joint statistics of covariates $\bx$ and $y$ in the population from which our samples are drawn. This choice is constrained by our desire to do all integrations over $\btheta$ analytically, to avoid approximations and overfitting problems caused by point-estimation. One is then naturally led to  class-specific Gaussian covariate distributions:
\begin{eqnarray}
p(\bx,y|\btheta)= p(y)p(\bx|y,\btheta),~~~~~~~p(\bx|y,\btheta)=\frac{\rme^{-\frac{1}{2}(\bx-\bmu_y)\cdot\bLambda_y(\bx-\bmu_y)}}{\sqrt{(2\pi)^d/{\rm Det}\bLambda_y}}
\label{eq:parametrisation}
\end{eqnarray} 
Thus the parameters to be integrated over are $\btheta=\{\bmu_y,\bLambda_y,~y=1\ldots C\}$, i.e. the class-specific means and covariate matrices. 

\subsection{Integrals to be computed}

In both the inference formula $p(y_0|\bx_0,\Data,H)$ (\ref{eq:inference_formula}) and in the expression for $\hat{H}$ (\ref{eq:maximise_evidence}), the relevant integral we need to do analytically is the one in 
\begin{eqnarray}
\Omega(H,n,\Data)&=& -\log \int\!\rmd\btheta ~p(\btheta|H)\prod_{i=1}^n p(\bx_i,y_i|\btheta,H)
\end{eqnarray}
In the case where we require $\Omega(H,n+1,\Data)$, when evaluating the numerator and the denominator of (\ref{eq:inference_formula}), we simply replace $\prod_{i=1}^n$ by $\prod_{i=0}^n$, so
that
\begin{eqnarray}
p(y_0|\bx_0,\Data)= \frac{\rme^{-\Omega(H,n+1,\Data)}}{\sum_{z=1}^C \rme^{-\Omega(H,n+1,\Data)}|_{y_0=z}}
\label{eq:prediction}
~~~~~~~~
\hat{H}= {\rm argmin}_H \Omega(H,n,\Data)
\end{eqnarray}
Working out $\Omega(H,n,\Data)$ for the  parametrization (\ref{eq:parametrisation}) gives:
\begin{eqnarray}
\Omega(H,n,\Data)&=& \frac{1}{2}nd\log(2\pi)-\sum_{i=1}^n \log p_{y_i}
\nonumber\\
&&
\hspace*{-5mm}
 -\log
\int\!\Big[\prod_{z=1}^C \rmd\bmu_z\rmd\bLambda_z ~p_z(\bmu_z,\bLambda_z)\Big] 
\Big[\prod_{i=1}^n ( {\rm Det}\bLambda_{y_i})^{\frac{1}{2}}\Big]
\rme^{-\frac{1}{2}\sum_{i=1}^n (\bx_i-\bmu_{y_i})\cdot\bLambda_{y_i}(\bx_i-\bmu_{y_i})}~~~
\end{eqnarray}
where $p(y_i) = p_{y_i}$ is the prior probability of a sample belonging to class $y_i$.
To simplify this expression we define the data-dependent index sets $I_z=\{i|~y_i=z\}$, each of size $n_z=|I_z|=\sum_{i=1}^n \delta_{z,y_i}$. We also introduce empirical covariate averages and correlations, with $\bx_i=(x_{i1},\ldots,x_{id})$:
\begin{eqnarray}
\hat{X}^z_{\mu}=\frac{1}{n_z}\sum_{i\in I_z} x_{i\mu},~~~~~~\hat{C}^z_{\mu\nu}=\frac{1}{n_z}\sum_{i\in I_z} (x_{i\mu}-\hat{X}^z_\mu) (x_{i\nu}-\hat{X}^z_\nu)
 \end{eqnarray}
Upon defining the vector $\hat{\bX}_z=(\hat{X}^z_1,\ldots,\hat{X}^z_d$), and the $d\!\times\!d$  matrix $\hat{\bC}_{\!z}=\{\hat{C}^z_{\mu\nu}\}$, we can then write the relevant integrals after some simple rearrangements in the form
\begin{eqnarray}
\Omega(H,n,\Data)\!\!&=& \! \! \frac{1}{2}nd\log(2\pi)-\sum_{z=1}^C n_z \log p_z 
\!
-\log \!
\int\!\!\Big[\!\prod_{z=1}^C \rmd\bmu_z\rmd\bLambda_z ~p_z(\bmu_z,\bLambda_z)({\rm Det}\bLambda_z)^{\frac{n_z}{2}}
\rme^{-\frac{1}{2} n_z \bmu_z\cdot\bLambda_{z}\bmu_{z}}\!\Big] 
\nonumber
\\&&\hspace*{70mm} \times~
\rme^{\sum_{z=1}^C \bmu_z\cdot\bLambda_{z}\sum_{i\in I_z}\bx_i-\frac{1}{2}\sum_{z=1}^C \sum_{i\in I_z} \bx_i\cdot\bLambda_{z}\bx_i}
\nonumber
\\[-1mm]
&=&  \frac{1}{2}nd\log(2\pi)-\sum_{z=1}^C n_z \log p_z 
\nonumber
\\[-1mm]&&
-\sum_{z=1}^C \log 
\int\!\rmd\bmu\rmd\bLambda ~p_z(\bmu\!+\!\hat{\bX}_z,\bLambda)({\rm Det}\bLambda)^{\frac{1}{2}n_z}
\rme^{-\frac{1}{2}  n_z\bmu\cdot\bLambda \bmu
-\frac{1}{2}  n_z{\rm Tr}(\hat{\bC}_z\bLambda)
}
\label{eq:Omega}
\end{eqnarray}
To proceed it is essential that we compute $\Omega(H,n,\Data)$  analytically, for arbitrary $\hat{\bX}\in\R^d$ and arbitrary positive definite symmetric matrices $\hat{\bC}$. This will constrain the choice of our priors $p_z(\bmu,\bLambda)$ for the covariate averages and correlations in outcome class $z$.  All required integrals are of the following form, with $\bLambda$ limited to the subset $\Xi^d$ of symmetric positive definite matrices:
\begin{eqnarray}
\Psi_z(H,n,\Data)&=& \int_{\R^d}\!\rmd\bmu\int_{\Xi^d}\!\rmd\bLambda ~p_z(\bmu\!+\!\hat{\bX}_z|\bLambda)p_z(\bLambda)({\rm Det}\bLambda)^{\frac{1}{2}n_z}
\rme^{-\frac{1}{2} n_z\bmu\cdot\bLambda \bmu
-\frac{1}{2} n_z{\rm Tr}(\hat{\bC}_z\bLambda)
}
\label{eq:Psi_z}
\end{eqnarray}
We will drop the indications of the sets over which the integrals are done, when these are clear from the context. 
The tricky integral is that over the inverse covariance matrices $\bLambda$. 
The choice in \cite{shalabi2016bayesian} corresponded to $p_z(\bmu,\bLambda)\propto \rme^{-\frac{1}{2}\bmu^2/\beta_z^2}\delta[\bLambda-\one/\alpha_z^2]$, which implied assuming uncorrelated covariates within each class. Here we want to allow for arbitrary class-specific covariate correlations.

\subsection{Priors for class-specific means and covariance matrices}

The integrals over $\bmu$ and $\bLambda$ can be done in either order.  We start with the integral over $\bmu$.
In contrast to most studies, we replace the conjugate prior for the unknown mean vector by a multivariate Gaussian with an as yet arbitrary precision matrix $\pra$. This should allow us to cover a larger parameter space than the conjugate prior (which has $\pr^{-1}$ as its covariance matrix):
\begin{eqnarray}
p_z(\bmu|\bA)&=& (2\pi)^{-\frac{d}{2}}\sqrt{{\rm Det}\bA_z}~\rme^{-\frac{1}{2}\bmu\cdot\bA_z\bmu}
\end{eqnarray}
Insertion into (\ref{eq:Psi_z}) gives
\begin{eqnarray}
\Psi_z&=&(2\pi)^{-\frac{d}{2}}\!\! \int\!\rmd\bLambda ~p_z(\bLambda)\rme^{-\frac{1}{2} n_z{\rm Tr}(\hat{\bC}_z\bLambda)-\frac{1}{2}\hat{\bX}_z\cdot\bA_z\hat{\bX}_z}
\Big[
{\rm Det}(\bLambda^{n_z})
{\rm Det}\bA_z\Big]^{\frac{1}{2}}
\nonumber
\\
&&\hspace*{30mm} \times
\int\!\rmd\bmu~
\rme^{-\frac{1}{2} \bmu\cdot(n_z\bLambda+\bA_z) \bmu
-\bmu\cdot\bA_z\hat{\bX}_z}
\nonumber
\\
&=& \int\!\rmd\bLambda ~p_z(\bLambda)\rme^{-\frac{1}{2} n_z{\rm Tr}(\hat{\bC}_z\bLambda)}
\Big[
\frac{{\rm Det}(\bLambda^{n_z}) {\rm Det}\bA_z}{{\rm Det}(n_z\bLambda\!+\!\bA_z)}\Big]^{\frac{1}{2}}
\rme^{
\frac{1}{2}\bX\cdot
\bA_z (n_z\bLambda+\bA_z)^{-1}\bA_z\hat{\bX}_z-\frac{1}{2}\hat{\bX}_z\cdot\bA_z\hat{\bX}_z}
\nonumber
\\
&=& \int\!\rmd\bLambda ~p_z(\bLambda)\rme^{-\frac{1}{2} n_z{\rm Tr}(\hat{\bC}_z\bLambda)}
\Big[{\rm Det}(n_z\bLambda^{1-n_z}\bA_z^{-1}+\bLambda^{-n_z})
\Big]^{-\frac{1}{2}}
\rme^{
-\frac{1}{2}\hat{\bX}\cdot
 [(n_z\bLambda)^{-1}+(\bA_z)^{-1}]^{-1}\hat{\bX}_z}
 ~~~~~
\label{eq:tricky_integral}
\end{eqnarray}
Our present more general assumptions lead to calculations that differ from the earlier work of e.g. \cite{keehn1965note,brown1999discrimination,srivastava2007bayesian}\footnote{Alternative analytically tractable priors are the transformation-invariant Jeffrey's or Reference priors, which are derived from information-theoretic arguments \cite{berger1992development}.  There the calculation of the predictive probability is simpler, but the sample covariance matrix is not regularized. This causes problems when  $n < d$, where the sample covariance matrices would become singular and the predictive probability would cease to be well defined.}. 
Our next question is for which choice(s) of $\bA_z$ we can do also the integrals over $\bLambda$ in (\ref{eq:tricky_integral}) analytically. Expression (\ref{eq:tricky_integral}), in line with \cite{keehn1965note,brown1999discrimination,srivastava2007bayesian}, suggests using for the
measure $p_z(\bLambda)$ over all positive definite matrices $\bLambda\in\Xi^d$ 
a Wishart distribution,  which is of the form
\begin{eqnarray}
p(\bLambda)&=& \frac{({\rm Det}\bLambda)^{(r-d-1)/2}}{2^{rd/2}\Gamma_{d}(\frac{r}{2})({\rm Det}\bS)^{r/2}}\rme^{-\frac{1}{2}{\rm Tr}(\bS^{-1}\bLambda)}
\label{eq:Wishart}
\end{eqnarray}
Here $r> d-1$,  $\bS$ is a positive definite and symmetric $d\times d$ matrix, and $\Gamma_p(x)$ is the multivariate Gamma function which is expressed in terms of the ordinary Gamma function via:
\begin{eqnarray}
\Gamma_p\Big(\frac{r}{2} \Big)&=& \pi^{p(p-1)/4}\prod_{j=1}^p \Gamma   \Big(\frac{r}{2}-\frac{j\!-\!1}{2}   \Big)
\end{eqnarray}
The choice (\ref{eq:Wishart})  is motivated solely by analytic tractability. 
However, since the prior domain is the space of all positive definite matrices, we are assured that upon using (\ref{eq:Wishart}) our posterior will be consistent. 
Distribution (\ref{eq:Wishart}) implies stating that $\bLambda$ is the empirical precision matrix of a set of $r$ i.i.d. random zero-average Gaussian vectors, with covariance matrix $\bS$. Since (\ref{eq:Wishart}) is normalised, for any $\bS$, we can use it to do all integrals of the following form analytically:
\begin{eqnarray}
\int_{\Xi^d}\rmd\bLambda ~({\rm Det}\bLambda)^{(r-d-1)/2}~\rme^{-\frac{1}{2}{\rm Tr}(\bS^{-1}\bLambda)}
&=& 
2^{rd/2}\Gamma_{d}   \Big(\frac{r}{2}   \Big)({\rm Det}\bS)^{r/2} 
\label{eq:doable_form}
\end{eqnarray}
In order for (\ref{eq:tricky_integral}) to acquire the form  (\ref{eq:doable_form}), we need a choice for $\bA_z$ such that the following holds, for some $\gamma_0,\gamma_1\in\R$:
$ [(n_z\bLambda)^{-1}\!+(\bA_z)^{-1}]^{-1}=\gamma_{1z}\bLambda+\gamma_{0z}\one$.
 Rewriting this condition gives:
 \begin{eqnarray}
\bA_z(\gamma_{0z},\gamma_1)&=&[(\gamma_{1z}\bLambda+\gamma_{0z}\one)^{-1}- (n_z\bLambda)^{-1}]^{-1}
\label{eq:canonical_A}
 \end{eqnarray}
 Clearly $\bA_z(\gamma_{0z},\gamma_1)$ has the same eigenvectors as $\bLambda$. Each eigenvalue $\lambda$ of $\bLambda$ would thus give a corresponding eigenvalue $a(\lambda,z)$ for $\bA_z(\gamma_{0z},\gamma_1)$:
  \begin{eqnarray}
a(\lambda,z)&=&\frac{n_z\lambda(\gamma_{1z}\lambda+\gamma_{0z})}
{(n_z-\gamma_{1z})\lambda-\gamma_{0z}}
 \end{eqnarray}
 We note that the zeros of $a(\lambda,z)$ occur at $\lambda\in\{-\gamma_{0z}/\gamma_{1z},0\}$, and that 
 \begin{eqnarray}
 \lambda\to 0:~~a(\lambda,z)= -n_z\lambda+\order(\lambda^2),~~~~~~
 \lambda\to\infty:~~a(\lambda,z)\approx \frac{n_z\gamma_{1z}\lambda}{n_z-\gamma_{1z}}
 \end{eqnarray}
The natural choice is to take
\begin{eqnarray}
\gamma_{1z}\in(0,n_z],~~~~~~
\gamma_{0z}\in(-\lambda_{\rm min}\gamma_{1z},\lambda_{\rm min}(n_z\!-\!\gamma_{1z}))
\end{eqnarray}
This ensures that always $a(\lambda,z)>0$. We expect that $\lambda_{\rm min}$ will increase monotonically with $r-d$. 
Upon making the choice (\ref{eq:canonical_A}) and using (\ref{eq:doable_form}), we obtain for the integral (\ref{eq:tricky_integral}):
\begin{eqnarray}
\Psi_z
&=& \rme^{-\frac{1}{2}\gamma_{0z}\hat{\bX}_z^2}\int\!\rmd\bLambda ~p_z(\bLambda)
\frac{\rme^{-\frac{1}{2} n_z{\rm Tr}(\hat{\bC}_z\bLambda)-\frac{1}{2}\gamma_{1z}\hat{\bX}_z\cdot\bLambda\hat{\bX}_z}}{
\sqrt{{\rm Det}[n_z\bLambda^{1-n_z}(\gamma_{1z}\bLambda+\gamma_{0z}\one)^{-1}]}}
\label{eq:tricky_integral_withA}
\end{eqnarray}
We conclude that we can evaluate (\ref{eq:tricky_integral_withA}) analytically, using (\ref{eq:doable_form}), provided we choose for $p_z(\bLambda)$ the Wishart measure, and with either $\gamma_{0z}\to 0$ and $\gamma_{1z}\in(0,n_z)$, or with $\gamma_{1z}\to 0$ and $\gamma_{0z}\in(0,n_z\lambda_{\rm min})$. 
Alternative choices for $(\gamma_{0z},\gamma_{1z})$ would lead to more complicated integrals than the Wishart one.

The two remaining analytically integrable candidate model branches imply the following choices for the inverse correlation matrix $\bA_z$ of the prior $p_z(\mu|\bA_z)$ for the class centres:
\begin{eqnarray}
\gamma_{0z}=0:~~~\bA_z=\frac{n_z\gamma_{1z}}{n_z\!-\!\gamma_{1z}}\bLambda,~~~~~~~~~\gamma_{1z}=0:~~~\bA_z=\Big[\gamma_{0z}^{-1}\one-(n_z\bLambda)^{-1}\Big]^{-1}
\end{eqnarray}
Note that the case $\bA_z\to\bnull$, a non-informative prior for class means as in \cite{brown1999discrimination}, corresponds to $(\gamma_{0z},\gamma_{1z})=(0,0)$.
However, the two limits $\gamma_{0z}\to 0$ and $\gamma_{1z}\to 0$ will generally not commute, which 
can be inferred from working out (\ref{eq:tricky_integral_withA}) for the two special cases $\gamma_{0z}=0$ and $\gamma_{1z}=0$:
\begin{eqnarray}
\gamma_{0z}=0: && 
\Psi_z
=\Big(\frac{\gamma_{1z}}{n_z}\Big)^{\!\frac{d}{2}} \int\!\rmd\bLambda ~p_z(\bLambda)[{\rm Det}(\bLambda)]^{\frac{n_z}{2}}
\rme^{-\frac{1}{2} n_z{\rm Tr}(\hat{\bC}_z\bLambda)-\frac{1}{2}\gamma_{1z}\hat{\bX}_z\cdot\bLambda\hat{\bX}_z}
\label{eq:tricky_integral_gamma0zero}
\\
\gamma_{1z}=0: && 
\Psi_z
= \Big(\frac{\gamma_{0z}}{n_z}\Big)^{\!\frac{d}{2}} \rme^{-\frac{1}{2}\gamma_{0z}\hat{\bX}_z^2}\int\!\rmd\bLambda ~p_z(\bLambda)
[{\rm Det}(\bLambda)]^{\frac{n_z-1}{2}}
\rme^{-\frac{1}{2} n_z{\rm Tr}(\hat{\bC}_z\bLambda)}
\label{eq:tricky_integral_gamma1zero}
\end{eqnarray}
This non-uniqueness of the limit $\bA_z\to\bnull$ is a consequence of having done the integral over $\bLambda$ first.

\section{The integrable model branches}
\label{sec:hyper}


\subsection{The case $\gamma_{0z}=0$: model A}

We now choose $\gamma_{0z}=0$, and substitute for each $z=1\ldots C$ the Wishart distribution (\ref{eq:Wishart}) into (\ref{eq:tricky_integral_withA}), with seed matrix $\bS=k_z\one$. This choice is named Quadratic Bayes in \cite{brown1999discrimination}. We also define the $p\!\times\! p$ matrix $\hat{\bM}_z$ with entries $\hat{M}^z_{\mu\nu}=X^z_\mu X^z_\nu$. The result of working out (\ref{eq:tricky_integral_withA}) is, using (\ref{eq:doable_form}):
\begin{eqnarray}
\Psi_z
&=&
\Big(\frac{2^{n_z}\gamma_{1z}}{n_z k_z^{r_z}}\Big)^{\!\frac{d}{2}}\frac{\Gamma_d(\frac{r_z+n_z}{2})}{\Gamma_d(\frac{r_z}{2})}
[{\rm Det}(n_z\hat{\bC}_z\!+\!\gamma_{1z}\hat{\bM}_z\!+\!k_z^{-1}\one)]^{-(r_z+n_z)/2} 
\end{eqnarray}
This, in turn, allows us to evaluate (\ref{eq:Omega}):
\begin{eqnarray}
\Omega(H,n,\Data)&=&  \frac{1}{2}nd\log(\pi)-\sum_{z=1}^C n_z \log p_z -\frac{1}{2}d\sum_{z=1}^C \Big[\log(\gamma_{1z}/n_z)\!-\!r_z\log k_z\Big]
-\sum_{z=1}^C \log \Big[\frac{\Gamma_d(\frac{r_z+n_z}{2})}{\Gamma_d(\frac{r_z}{2})}\Big]
\nonumber
\\&&
+~\frac{1}{2}\sum_{z=1}^C (r_z\!+\!n_z)\log 
{\rm Det}(n_z\hat{\bC}_z\!+\!\gamma_{1z}\hat{\bM}_z\!+\!k_z^{-1}\one)
\label{eq:Omega_gamma0_zero}
\end{eqnarray}
The hyperparameters of our problem are $\{p_z,\gamma_{1z},r_z,k_z\}$, for $z=1\ldots C$. If we choose flat hyper-priors, to close the Bayesian inference hierarchy, their optimal values are obtained by minimizing (\ref{eq:Omega_gamma0_zero}), subject to the constraints $\sum_{z=1}^C p_z=1$, $p_z\geq 0$, $r_z\geq d$, $\gamma_{1z}\in[0,n_z]$, and $k_z>0$. 
We now work out the relevant extremization equations, using the general identity $\partial_x\log {\rm Det}\bQ={\rm Tr}(\bQ^{-1}\partial_x \bQ)$:
\begin{itemize}
\item Minimization over $p_z$: $~p_z=n_z/n$.
 \item Minimization over $k_z$:
\begin{eqnarray}
k_z=0&~{\rm or}~ & 
r_z=n_z\Big[\frac{dk_z}{{\rm Tr}[
(n_z\hat{\bC}_z\!+\!\gamma_{1z}\hat{\bM}_z\!+\!k_z^{-1}\one)^{-1}}-1\Big]^{-1}
\end{eqnarray}
 \item Minimization over $r_z$, using the digamma  function $\psi(x)=\frac{\rmd}{\rmd x}\log\Gamma(x)$:
 \begin{eqnarray}
r_z=d &~{\rm or}~&  
 \log k_z
=\frac{1}{d}\sum_{j=1}^d \Big[
\psi(\frac{r_z\!+\!n_z\!-\!j\!+\!1}{2})
-\psi(\frac{r_z\!-\!j\!+\!1}{2})\Big]
\nonumber
\\[-1mm]&&
\hspace*{30mm}
-\frac{1}{d}\log 
{\rm Det}(n_z\hat{\bC}_z\!+\!\gamma_{1z}\hat{\bM}_z\!+\!k_z^{-1}\one)
\Big]
\\[-8mm]&&\nonumber
\end{eqnarray}
 \item Minimization over $\gamma_{1z}$:
\begin{eqnarray}
\gamma_{1z}\in\{0,n_z\} &~{\rm or}~& 
\gamma_{1z}
=\frac{1} {r_z\!+\!n_z}\Big[\frac{1}{d}{\rm Tr}[
(n_z\hat{\bC}_z\!+\!\gamma_{1z}\hat{\bM}_z\!+\!k_z^{-1}\one)^{-1}\hat{\bM}_z]\Big]^{-1}
\end{eqnarray}
\end{itemize}
In addition we still need to satisfy the inequalities $r_z\geq d$, $\gamma_{1z}\in[0,n_z]$, and $k_z>0$. 

We observe in the above results that, unless we choose $\gamma_{1z}\in\{0,n_z\}$, i.e. $\bA=\bnull$ or $\bA^{-1}=\bnull$, we would during any iterative algorithmic solution of our order parameter equations have to diagonalize a $d\times d$ matrix at each iteration step. 
This would be prohibitively slow, even with the most efficient numerical diagonalization methods. 
Since $\gamma_{1z}=n_z$ implies that the prior $p_z(\bmu|\bA)$ forces all class centres to be in the origin, we will be left for the current model branch only with the option $\gamma_{1z}\to 0$, corresponding to a flat prior for the class centres. 
We thereby arrive at the Quadratic Bayes model of \cite{brown1999discrimination}, with hyperparameter formulae based on evidence maximization. 

\subsection{The case $\gamma_{1z}=0$: model B}

We next inspect the alternative model branch by choosing $\gamma_{1z}=0$, again substituting for each $z=1\ldots C$ the Wishart distribution (\ref{eq:Wishart}) into (\ref{eq:tricky_integral_withA}) with seed matrix $\bS=k_z\one$. 
The result is:
\begin{eqnarray}
\Psi_z
&=&
\Big(\frac{2^{n_z-1}\gamma_{0z}}{n_z k_z^{r_z}}\Big)^{\!\frac{d}{2}}\frac{\Gamma_d(\frac{r_z+n_z-1}{2})}{\Gamma_d(\frac{r_z}{2})}
[{\rm Det}(n_z\hat{\bC}_z\!+\!k_z^{-1}\one)]^{-(r_z+n_z-1)/2} \rme^{-\frac{1}{2}\gamma_{0z}\hat{\bX}^2_z}
\end{eqnarray}
For the quantity (\ref{eq:Omega}) we thereby find:
\begin{eqnarray}
\Omega(H,n,\Data)&=&  \frac{1}{2}nd\log(\pi)+\frac{1}{2}dC\log 2-\sum_{z=1}^C n_z \log p_z 
-\frac{1}{2}d\sum_{z=1}^C \Big[\log(\frac{\gamma_{0z}}{n_z})\!-\!r_z\log k_z\Big]
\label{eq:Omega_gamma1_zero}
\\&&
-\sum_{z=1}^C \log \Big[\frac{\Gamma_d(\frac{r_z+n_z-1}{2})}{\Gamma_d(\frac{r_z}{2})}\Big]
+\frac{1}{2}\sum_{z=1}^C \gamma_{0z}\hat{\bX}^2_z
+~\frac{1}{2}\sum_{z=1}^C (r_z\!+\!n_z\!-\!1)\log 
{\rm Det}(n_z\hat{\bC}_z\!+\!k_z^{-1}\one)
\nonumber
\end{eqnarray}
If as before we choose flat hyper-priors, 
the Bayes-optimal hyperparameters  $\{p_z,\gamma_{1z},r_z,k_z\}$, for $z=1\ldots C$ are found by maximizing the evidence (\ref{eq:Omega_gamma1_zero}), subject to the constraints $\sum_{z=1}^C p_z=1$, $p_z\geq 0$, $r_z\geq d$, $\gamma_{0z}\geq 0$, and $k_z>0$. For the present model branch B, differentiation gives
\begin{itemize}
\item Minimization over $p_z$: $~p_z=n_z/n$.
 \item Minimization over $k_z$:
\begin{eqnarray}
k_z=0&~{\rm or}~ & 
r_z=(n_z\!-\!1)\Big[\frac{dk_z}{{\rm Tr}[
(n_z\hat{\bC}_z\!+\!k_z^{-1}\one)^{-1}}-1\Big]^{-1}
\end{eqnarray}
 \item Minimization over $r_z$:
 \begin{eqnarray}
r_z=d &~{\rm or}~&  
 \log k_z
=\frac{1}{d}\sum_{j=1}^d \Big[
\psi(\frac{r_z\!+\!n_z\!-\!j}{2})\!
-\psi(\frac{r_z\!-\!j\!+\!1}{2})\Big]
-\frac{1}{d}\log 
{\rm Det}(n_z\hat{\bC}_z\!+\!k_z^{-1}\one)~~~~
\end{eqnarray}
 \item Minimization over $\gamma_{0z}$:
$~\gamma_{0z}=d/\hat{\bX}^2_z$.
\end{itemize}
In addition we still need to satisfy the inequalities $r_z\geq d$ and $k_z>0$. 
In contrast to the first integrable model branch A, here we are able to optimise over $\gamma_{0z}$ without problems, and 
the resulting model B is distinct from the Quadratic Bayes classifier of  \cite{brown1999discrimination}.

\subsection{Comparison of the two integrable model branches}

Our initial family of models was parametrized by $(\gamma_{0z},\gamma_{1z})$. We then found that the following  two branches  are analytically integrable, using Wishart priors for class-specific precision matrices: 
\begin{eqnarray}
{\rm A}: && (\gamma_{0z},\gamma_{1z})=(0,\hat{\gamma}_{1z})~~{\rm with}~~ \hat{\gamma}_{1z}\to 0
\\
{\rm B}: && (\gamma_{0z},\gamma_{1z})=(\hat{\gamma}_{0z},0)~~{\rm with}~~ \hat{\gamma}_{0z}\to d/\hat{\bX}_z^2
\end{eqnarray}
Where conventional methods tend to determine hyperparameters via cross-validation, which is computationally expensive, here we optimize hyperparameters via evidence maximization.  As expected, both models give $p_z=n_z/n$. The hyperparameters $(k_z,r_z)$ are
 to be solved from the following equations, in which $\varrho_z(\xi)$ denotes the eigenvalue distribution of $\hat{\bC}_z$:
\begin{eqnarray}
{\rm A}: &&  k_z=0~~~{\rm or}~~~  
r_z=n_z\Big[\frac{1}{\int\!\rmd\xi~\varrho_z(\xi)
(n_zk_z\xi\!+\!1)^{-1}}-1\Big]^{-1}
\\
&&
r_z=d ~~~{\rm or}~~~
\frac{1}{d}\sum_{j=1}^d \Big[
\psi(\frac{r_z\!+\!n_z\!-\!j\!+\!1}{2})
-\psi(\frac{r_z\!-\!j\!+\!1}{2})\Big]
=\int\!\rmd\xi~\varrho_z(\xi)\log 
(n_zk_z\xi\!+\!1)
\label{eq:modelA_reqn}
\\
{\rm B}: && k_z=0~~~{\rm or}~~~ 
r_z=(n_z\!-\!1)\Big[\frac{1}{\int\!\rmd\xi~\varrho_z(\xi)
(n_zk_z\xi\!+\!1)^{-1}}-1\Big]^{-1}
\\
&&
r_z=d ~~{\rm or}~~  
\frac{1}{d}\sum_{j=1}^d \Big[
\psi(\frac{r_z\!+\!n_z\!-\!j}{2})
-\psi(\frac{r_z\!-\!j\!+\!1}{2})\Big]
=\int\!\rmd\xi~\varrho_z(\xi)\log 
(n_zk_z\xi\!+\!1)
\end{eqnarray}
We see that the equations for $(k_z,r_z)$ of models A and B differ only in having the replacement $n_z\to n_z\!-\!1$ in certain places. 
Hence we will have $(k_z^A,r_z^A)=(k_z^B,r_z^B)+\order(n_z^{-1})$.

\subsection{Expressions for the predictive probability}
\label{sec:predictive}

We can now proceed to work out formula (\ref{eq:prediction}) for the class prediction probabilities $p(y_0|\bx_0,\Data)$.
This requires making the replacements $n\to n\!+\!1$ and $n_z\to n_z\!+\!\delta_{y_0,z}$ (taking care not to change the hyperparameter equations), and 
\begin{eqnarray}
\hat{X}_\mu^{z} &\to &   \hat{X}_\mu^{z}  +\frac{\delta_{z y_0}}{n_{y_0}\!+\!1}\Big( x_{0\mu}
-\hat{X}_\mu^{z} \Big)\\
\nonumber
\\
\hat{C}_{\mu\nu}^z & \to & \hat{C}_{\mu\nu}^z+\frac{\delta_{z y_0}}{n_{y_0}\!+\!1}
\Big[
\frac{n_{y_0}}{n_{y_0}\!+\!1}
(x_{0\mu}\!-\!\hat{X}_\mu^{z})
(x_{0\nu}\!-\!\hat{X}_\nu^{z})
-\hat{C}_{\mu\nu}^{y_0}
\Big]
\end{eqnarray}
We introduce the $d\times d$ matrix $\bM_{y_0}$, with entries $M_{\mu\nu}^{y_0}=(x_{0\mu}\!-\!\hat{X}_\mu^{y_0})(x_{0\nu}\!-\!\hat{X}_\nu^{y_0})$. 
This leads us to 
\begin{eqnarray}
\Omega(H,n\!+\!1,\Data)-\Omega(H,n,\Data) 
&=& 
 \frac{1}{2}d\log(\pi) - \log p_{y_0}
-\frac{1}{2}d \log(\frac{n_{y_{0}}}{n_{y_0}+1})
- \log \Big[\frac{\Gamma_d(\frac{r_{y_0}+n_{y_0}}{2})}{\Gamma_d(\frac{r_{y_0}+n_{y_0}-1}{2})}\Big]
\nonumber
\\&&
+\frac{1}{2} \gamma_{0 y_0}\Big[
 \frac{2}{n_{y_0}\!+\!1}\hat{\bX}_{y_0}  \cdot( \bx_{0}\!-\!\hat{\bX}_{y_0} )
+\frac{1}{(n_{y_0}\!+\!1)^2}( \bx_{0}\!-\!\hat{\bX}_{y_0} )^2
\Big]
\\&&
+\frac{1}{2} \log 
{\rm Det}[\bXi_{y_0}]
+\frac{1}{2} (r_{y_0}\!+\!n_{y_0})\log {\rm Det}\Big[\one+\bXi_{y_0}^{-\frac{1}{2}}
\frac{n_{y_0}\bM_{y_0}}{n_{y_0}\!+\!1}\bXi_{y_0}^{-\frac{1}{2}}\Big]
\nonumber
\end{eqnarray}
with the short-hand $\bXi_z=n_{z}\hat{\bC}_{z}\!\!+\!k_{z}^{-1}\one$.  Note that $\bXi_z$ and $\hat{\bC}_{z}$  share the same eigenvectors. 
 The term with the Gamma functions can be simplified to:
\begin{eqnarray}
\frac{\Gamma_d(\frac{r_{y_0}+n_{y_0}}{2})}{\Gamma_d(\frac{r_{y_0}+n_{y_0}-1}{2})}&=& 
\prod_{j=1}^d \frac{\Gamma(\frac{r_{y_0}+n_{y_0}-j+1}{2})}{\Gamma(\frac{r_{y_0}+n_{y_0}-j}{2})}  = \frac{\Gamma(\frac{r_{y_0}+n_{y_0}}{2})}{\Gamma(\frac{r_{y_0}+n_{y_0}-d}{2})} \\
\nonumber
\end{eqnarray}
Since $\bM_{y_0}$ is proportional to a projection, the symmetric matrix $\bXi_{y_0}^{-\frac{1}{2}}\bM_{y_0}\bXi_{y_0}^{-\frac{1}{2}}$ has only one nonzero eigenvalue $\lambda_{y_0}$. 
This nontrivial eigenvalue can be computed via
\begin{eqnarray}
\lambda_{y_0}&=& {\rm Tr}[\bXi_{y_0}^{-\frac{1}{2}}\bM_{y_0}\bXi_{y_0}^{-\frac{1}{2}}]
~=~
\sum_{\mu\nu=1}^d (\bXi_{y_0}^{-1})_{\mu\nu}(x_{0\mu}\!-\!\hat{X}_\mu^{y_0})(x_{0\nu}\!-\!\hat{X}_\nu^{y_0})
\nonumber
\\&=& (\bx_0\!-\!\hat{\bX}_{y_0})\cdot\bXi_{y_0}^{-1}(\bx_0\!-\!\hat{\bX}_{y_0})
\end{eqnarray}
Hence
\begin{eqnarray}
\Omega(H,n\!+\!1,\Data)&=& \Omega(H,n,\Data)
+
 \frac{1}{2}d\log(\pi) - \log p_{y_0}
-\frac{1}{2}d \log(\frac{n_{y_{0}}}{n_{y_0}\!+\!1})
- \log \Big[\frac{\Gamma(\frac{r_{y_0}+n_{y_0}}{2})}{\Gamma(\frac{r_{y_0}+n_{y_0}-d}{2})}]
\nonumber
\\&&
+\frac{1}{2} \gamma_{0 y_0}\Big[
 \frac{2}{n_{y_0}\!+\!1}\hat{\bX}_{y_0}  \cdot( \bx_{0}\!-\!\hat{\bX}_{y_0} )
+\frac{1}{(n_{y_0}\!+\!1)^2}( \bx_{0}\!-\!\hat{\bX}_{y_0} )^2
\Big]
\nonumber
\\&&\hspace*{-5mm}
+\frac{1}{2} \log 
{\rm Det}[\bXi_{y_0}]
+\frac{1}{2} (r_{y_0}\!+\!n_{y_0})\log \Big[1+
\frac{n_{y_0}}{n_{y_0}\!+\!1}
(\bx_0\!-\!\hat{\bX}_{y_0})\cdot\bXi_{y_0}^{-1}(\bx_0\!-\!\hat{\bX}_{y_0})
\Big]~~~~
\end{eqnarray}
This then leads after some simple manipulations to the predictive probability for model B:
\begin{eqnarray}
p(y_0|\bx_0,\Data)&=&
\label{eq:prediction_B}
\\[0mm]
&&\hspace*{-22mm}
 \frac{W_{y_0}
\rme^{- \frac{\gamma_{0 y_0}}{2(n_{y_0}\!+\!1)}\Big[
 2\hat{\bX}_{y_0}  \cdot( \bx_{0}\!-\!\hat{\bX}_{y_0} )
+\frac{1}{n_{y_0}\!+\!1}( \bx_{0}\!-\!\hat{\bX}_{y_0} )^2
\Big]
}
\Big(1\!+\!\frac{n_{y_0}}{n_{y_0}\!+\!1}(\bx_0\!-\!\hat{\bX}_{y_0})\cdot\bXi_{y_0}^{-1}(\bx_0\!-\!\hat{\bX}_{y_0})\Big)^{-\frac{1}{2} (r_{y_0}\!+\!n_{y_0})}
}
{\sum_{z=1}^C
W_z
\rme^{-\frac{ \gamma_{0 z}}{2(n_{z}\!+\!1)}\Big[
 2\hat{\bX}_{z}  \cdot( \bx_{0}\!-\!\hat{\bX}_{z} )
+\frac{1}{n_{z}\!+\!1}( \bx_{0}\!-\!\hat{\bX}_{z} )^2
\Big]
}
\Big(1\!+\!\frac{n_{z}}{n_{z}\!+\!1}(\bx_0\!-\!\hat{\bX}_{z})\cdot\bXi_{y_0}^{-1}(\bx_0\!-\!\hat{\bX}_{z})\Big)^{-\frac{1}{2} (r_{z}\!+\!n_{z})}
}
\nonumber
\end{eqnarray}
with $p_z=n_z/n$, and
\begin{eqnarray}
W_z= p_{z}\big(\frac{n_{z}}{n_{z}\!+\!1}\big)^{\frac{d}{2}}\frac{\Gamma\big(\frac{r_{z}+n_{z}}{2}\big)}
{\Gamma\big(\frac{r_{z}+n_{z}-d}{2}\big)}[{\rm Det}\bXi_{z}]^{-\frac{1}{2}},~~~~~~\gamma_{0z}=d/\hat{\bX}^2_z,~~~~~~\bXi_{z}=n_{z}\hat{\bC}_{z}\!\!+\!k_{z}^{-1}\one
\end{eqnarray}
Upon repeating the same calculations for model A one finds that its predictive probability is obtained form expression (\ref{eq:prediction_B})
simply by setting $\gamma_{0 y_0}$ to zero (keeping in mind that for model A we would also insert into this formula distinct values for the optimal hyperparameters $k_z$ and $r_z$).

\section{Phenomenology of the classifiers}

\subsection{Hyperparameters: LOOCV versus evidence maximization}
\label{sec:behave}
\label{sec:optimal}

\begin{figure}[t]
\centering
\vspace*{-5mm}
\begin{center}
\includegraphics[width=8.5cm]{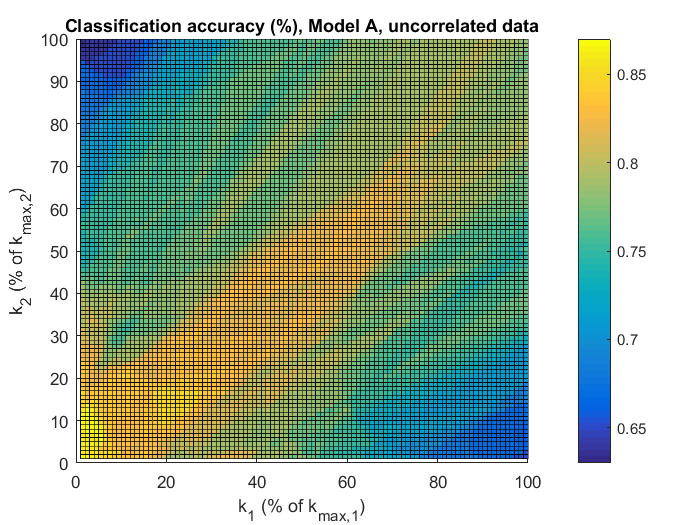} \includegraphics[width=8.5cm]{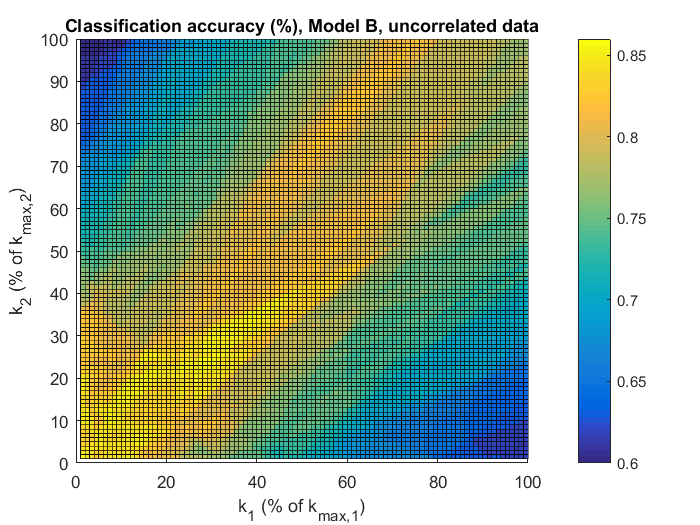} 
\end{center}
\vspace*{-4mm}
\caption{LOOCV classification accuracy in $(k_1,k_2)$  space for uncorrelated synthetic data, with class means $\bmu_1=(0,0,\ldots,0)$ and $\bmu_2=(2.5,0,\ldots,0)$, population covariance matrices  $\sig_1=\sig_2=\I$, and covariate dimension $d=50$. The hyperparameters $(r_1,r_2)$ for models A and B were determined via equation  (\ref{eq:modelA_reqn}). The results are based on a single data realisation. }
\label{fig:landscape1}
\end{figure}


\begin{figure}[t]
\centering
\vspace*{-5mm}
\begin{center}
\includegraphics[width=8.5cm]{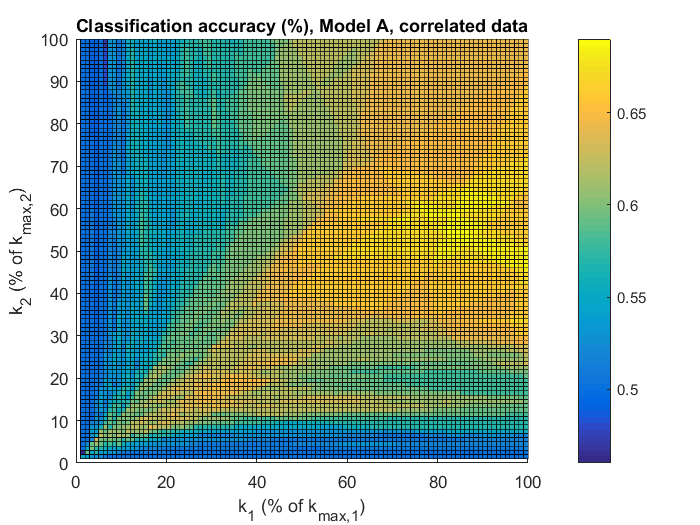} \includegraphics[width=8.5cm] {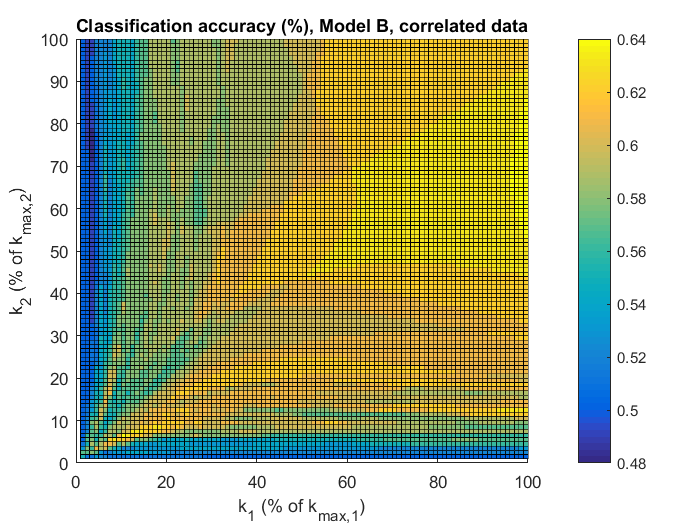} 
\end{center}
\vspace*{-4mm}
\caption{LOOCV classification accuracy in $(k_1,k_2)$  space for correlated synthetic data, with  class means $\bmu_1=(0,0,\ldots,0)$ and $\bmu_2=(2.5,0,\ldots,0)$,  population covariance matrices  $\sig_1=\sig_2=\sig$ of symmetric Toeplitz form with first row $(d,d-1,\ldots, 2,1)$,  and covariate dimension $d=50$. The hyperparameters $(r_1,r_2)$ for models A and B were determined via equation  (\ref{eq:modelA_reqn}). The results are based on a single data realisation. }
\label{fig:landscape2}
\end{figure}


The most commonly used measure for classification performance is the percentage of samples correctly predicted on unseen data (equivalently, the trace of the confusion matrix), and most Bayesian classification methods also use this measure as the optimization target for hyperparameters, via cross-validation. Instead, our method of hyperparameter optimization maximizes the evidence term in the Bayesian inference. In $k$-fold cross-validation one needs to diagonalize for each outcome class a $d\times d$ matrix $k$ times, whereas using the evidence maximization route one needs to diagonalize such matrices only once, giving a factor $k$ reduction in what for large $d$ is the dominant contribution to  the numerical demands.  Moreover, cross-validation introduces fluctuations into the hyperparameter computation (via the random separations into training and validation sets), whereas evidence maximization is strictly deterministic. 

The two routes, cross-validation versus evidence maximization, need not necessarily lead to coincident hyperparameter estimates. 
In order to investigate such possible differences 
we generated synthetic data-sets with equal class sizes $n_1=n_2=50$, and with input vectors of dimension $d=50$. Using a $100\times 100$ grid of values for the hyperparameters $k_1$ and $k_2$, with $k_z\in[0,k_{{\rm max}, z}]$, we calculated the leave-one-out cross-validation (LOOCV) estimator of the classification accuracy for unseen cases, for a single data realisation. The values of $(r_1,r_2)$ were determined via evidence maximization, using formula (\ref{eq:modelA_reqn}) (i.e. following model branch A, with the noninformative prior for the class means).  The value $k_{{\rm max},z}$ is either the upper limit defined by the condition $r_z>d-1$ (if such a limit exists, dependent on the data realisation), otherwise set numerically to a fixed large value. The location of the maximum of the resulting surface determines the LOOCV estimate of the optimal hyperparameters $(k_1,k_2)$, which can be compared to the optimized hyperparameters $(k_1,k_2)$  of the evidence maximization method. 

\begin{table}[h]
\centering
    \begin{tabular}{| c  || c   || c | c  |  }
 \hline
$(k_1,k_2)$  & $method$   & {\em model A}  & {\em model B}   \\ \hline  \hline
    {\bf Uncorrelated data\rule{0pt}{3ex}}    & Cross-validation  &    (1-5\%, 1-13\%)	& (1\%, 1\%)  \\ \hline
                  & Evidence maximization  &  (3\%, 2\%)  	& (2\%, 1\%)  \\ \hline \hline

    {\bf Correlated data\rule{0pt}{3ex}}    & Cross-validation   &  (82-87\%, 55-61\%) 	& (96-100\%, 54-92\%) \\ \hline
                  & Evidence maximization  &    (94\%, 95\%)	& (94\%, 94\%) \\ \hline
 \end{tabular}\vspace*{2mm}
\caption{\small Comparison of hyperparameter estimation using cross-validation and evidence maximization for correlated and uncorrelated data. Entries are the values of $(k_1,k_2)$, given as a percentage of each class $k_{max}$, corresponding to the maximum classification accuracy (within the granularity of our numerical experiments). A range of values is given when they all share the same classification accuracy.}
\label{tab:optimal}
\end{table}

Figure \ref{fig:landscape1} shows the resulting surface for uncorrelated data, i.e. $\sig_1=\sig_2=\I$. The comparison points from our evidence-based optimal hyperparameters $(k_1,k_2)$ are shown in table \ref{tab:optimal}. 
The small values for $(k_1,k_2)$ imply that the model correctly infers that the components of $\bx$ in each class are most likely uncorrelated. 
The same protocol was subsequently repeated for correlated data, using a Toeplitz covariance matrix, the results of which are shown in Figure \ref{fig:landscape2} and table \ref{tab:optimal}.
The larger values for $(k_1,k_2)$ imply that here the model correctly infers that the components of $\bx$ in each class are correlated. 
In both cases the differences between optimal hyperparameter values defined via LOOCV as opposed to evidence maximization are seen to be minor. 

\begin{table}[h]
\centering
    \begin{tabular}{| c  || c   || c | c  |  }
 \hline
\emph{classification accuracy (\%)}  & $method$   & {\em model A}  & {\em model B}   \\ \hline  \hline
    {\bf Uncorrelated data \rule{0pt}{3ex}}    & Cross-validation  &    87\% & 86\% \\ \hline
                  & Evidence maximization  &  86\% & 83\%  \\ \hline \hline

    {\bf Correlated data\rule{0pt}{3ex}}    & Cross-validation   &  69\% 	& 64\% \\ \hline
                  & Evidence maximization  &  64\% & 62\% \\ \hline
 \end{tabular}\vspace*{2mm}
\caption{\small Comparison of classification accuracy using cross-validation and evidence maximization methods for estimating hyperparameters using the same data as Figures \ref{fig:landscape1} and \ref{fig:landscape2}.  }
\label{tab:optimal_acc}
\end{table}

\subsection{Overfitting}

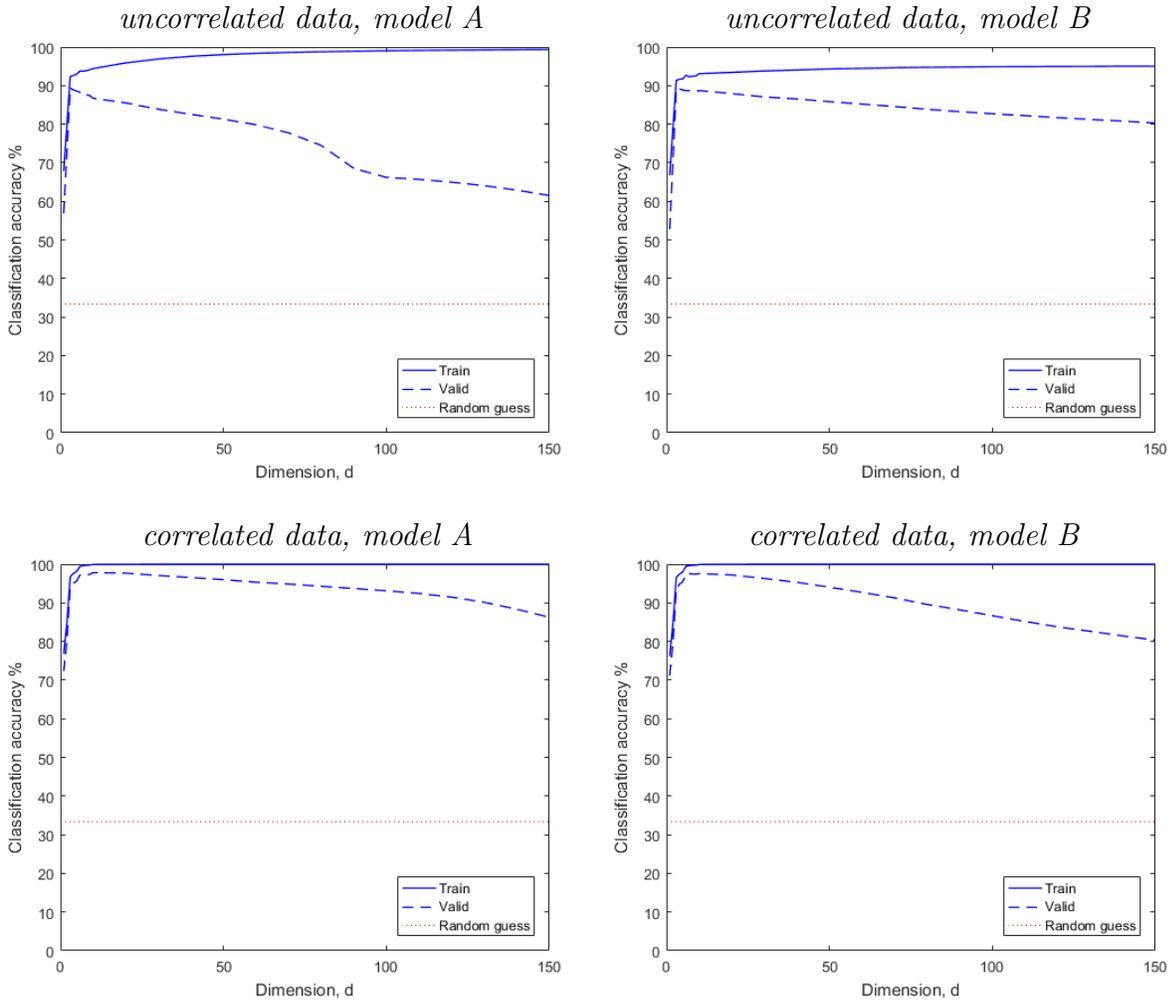
\begin{figure}[t]
\unitlength=0.62mm
\hspace*{0mm}
\begin{picture}(200,214)

\put(0,111){\includegraphics[width=135\unitlength]{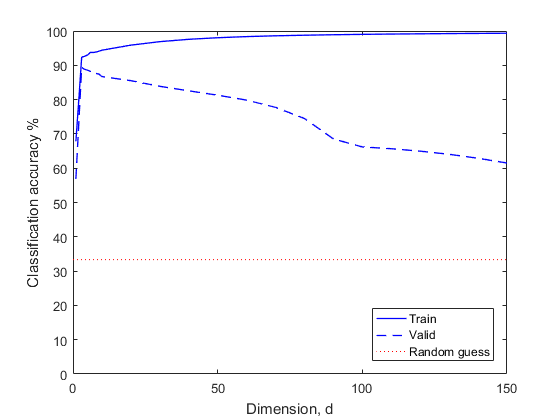}}
\put(30,209){\em uncorrelated data, model A}
\put(130,111){\includegraphics[width=135\unitlength]{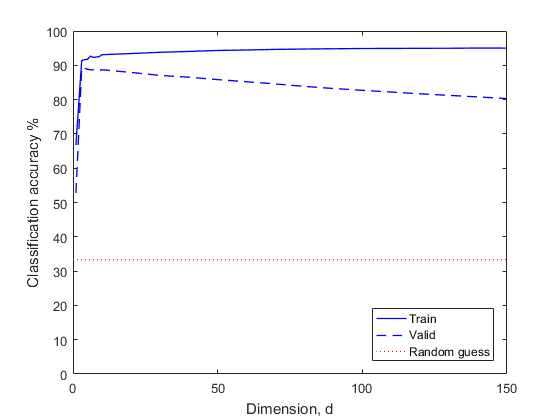}}
\put(160,209){\em uncorrelated data, model B}

\put(0,0){\includegraphics[width=135\unitlength]{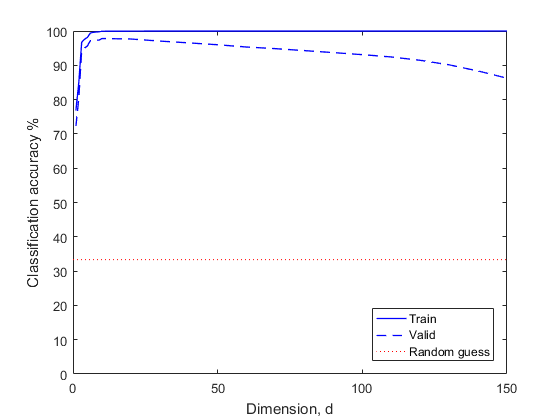}}
\put(35,98){\em correlated data, model A}
\put(130,0){\includegraphics[width=135\unitlength]{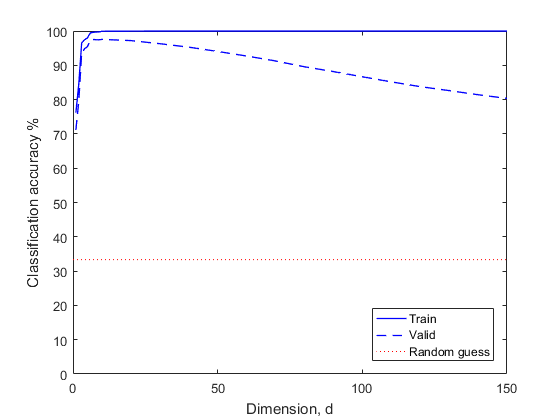}}
\put(165,98){\em correlated data, model B}
\end{picture}\vspace*{1mm}

    \caption{\small Overfitting in models A and B as measured via LOOCV. Top: uncorrelated data (case 1 in Table \ref{tab:syn}). Bottom: correlated data (case 8 in Table \ref{tab:syn}). 
   In all cases  $n_z=13$  for each of the three classes. Solid lines: classification accuracy on training samples; dashed lines: classification accuracy on validation  samples. Horizontal dotted line: baseline performance of a random guess classifier. }
\label{fig:over}
\end{figure}

Next we illustrate the degree of overfitting for models A and B, using examples of both correlated and uncorrelated synthetic data-sets. We sampled from the data described in Table \ref{tab:syn}, using case 1 (uncorrelated) and case 8 (correlated).
In all cases we chose $C=3$  classes of $n_z=13$ samples each, for a broad range of data dimensions $d$.  See the caption of Table \ref{tab:syn} for a full description of the statistical features of these synthetic data-sets.  
Measuring training and validation classification performance via LOOCV on these data resulted in Figure \ref{fig:over}, where each data-point is an average over $250$ simulation experiments. 
The degree of divergence between the training and validation curves (solid versus dashed)  is a direct measure  of the degree of overfitting. We observe that 
model B overfits less for uncorrelated data, and model A overfits less for correlated data. This pattern is also seen more generally in Table \ref{tab:syn_res}, for a broader range of synthetic data-sets. 
However, we note that all models still perform significantly above the random guess level on unseen data, even when $d \gg n_z$. For instance, for $d=150$ (corresponding to $d/n_z\approx 11.5$) the Bayesian models can still classify some 80\% of the unseen data correctly.

\section{Classification accuracy}
\label{sec:results}

We compare the classification accuracy of our Bayesian models A and B, with hyperparameters optimized by evidence maximization,  to other successful state-of-the-art generative classifiers from \cite{srivastava2007bayesian}. These include the distribution-based Bayesian classifier (BDA7), the Quadratic Bayes (QB) classifier \cite{brown1999discrimination}, and a non-Bayesian method, the so-called eigenvalue decomposition discriminant analysis (EDDA) as described in \cite{bensmail1996regularized}.  
All three use cross-validation for model selection and hyperparameter estimation. 
The classifiers (our models A and B and the three benchmark methods from \cite{srivastava2007bayesian}) are all tested on the same synthetic and real data-sets,  and following the definitions and protocols described in \cite{srivastava2007bayesian}, for a fair comparison. Model A differs from Quadratic Bayes \cite{brown1999discrimination} only in that our hyperparameters have been estimated using evidence maximization, as described in section \ref{sec:hyper}, rather than via cross-validation. Model A is seen in Table \ref{tab:syn_res}  to have lower error rates than Quadratic Bayes in the majority of the synthetic data-sets.  In contrast, model B is mathematically different from both model A and Quadratric Bayes.

\subsection{Synthetic data}

\begin{sidewaystable}
\centering
    \begin{tabular}{| c | c  | c | c | c | c | c |  }
 \hline
\rule{0pt}{3ex} & $\sig_1$  &  $\sig_2$  & $\sig_3$   & $\bmu_1$ & $\bmu_2$  & $\bmu_3$ \\ \hline  \hline
    {\bf  Case 1 \rule{0pt}{5ex} }    & $\I$  &  $\I$  &  $\I$	 &	$(0,0,\ldots,0)$  &  $(3,0,\ldots,0,0)$	& $(0,0,\ldots,0,3)$ \\ \hline

    {\bf Case 2  \rule{0pt}{5ex}}   & $\I$   &  $2 \I$  &  $3\I$	 &	$(0,0,\ldots,0)$ &  $(3,0,\ldots,0,0)$   &  $(0,0,\ldots,0,4)$	\\ \hline

    {\bf Case 3 \rule{0pt}{5ex}}   & $\Sigma_{ii}=\big(\frac{9(i-1)}{d-1} + 1\big)^2$   &  $\Sigma_{ii}=\big(\frac{9(i-1)}{d-1} + 1\big)^2$  &  $\Sigma_{ii}=\big(\frac{9(i-1)}{d-1} + 1\big)^2$ & $(0,0,\ldots,0)$ & $\mu_{2i}=2.5 \sqrt \frac{e_i}{d} \big(\frac{d-i}{\dhalf - 1}\big)$   &  $\mu_{3i}=(-1)^i \mu_{2i}$	\\ \hline

    {\bf Case 4  \rule{0pt}{5ex}}   & $\Sigma_{ii}=\big(\frac{9(i-1)}{d-1} + 1\big)^2$   &  $\Sigma_{ii}=\big(\frac{9(i-1)}{d-1} + 1\big)^2$  &  $\Sigma_{ii}=\big(\frac{9(i-1)}{d-1} + 1\big)^2$ & $(0,0,\ldots,0)$ & $\mu_{2i}=2.5 \sqrt \frac{e_i}{d} \big(\frac{i-1}{\dhalf - 1}\big)$   &  $\mu_{3i}=(-1)^i \mu_{2i}$	\\ \hline

    {\bf Case 5  \rule{0pt}{5ex}}   & $\Sigma_{ii}=\big(\frac{9(i-1)}{d-1} + 1\big)^2$   &  $\Sigma_{ii}=\big(\frac{9(d-i)}{d-1} + 1\big)^2$  &  $\Sigma_{ii}=\big(\frac{9(i-(\frac{d-1}{2}))}{d-1} + 1\big)^2$ & $(0,0,\ldots,0)$ & $(0,0,\ldots,0)$   &  $(0,0,\ldots,0)$  	\\   \hline

    {\bf Case 6  \rule{0pt}{5ex}}   & $\Sigma_{ii}=\big(\frac{9(i-1)}{d-1} + 1\big)^2$   &  $\Sigma_{ii}=\big(\frac{9(d-i)}{d-1} + 1\big)^2$  &  $\Sigma_{ii}=\big(\frac{9(i-(\frac{d-1}{2}))}{d-1} + 1\big)^2$ & $(0,0,\ldots,0)$ & $(\frac{14}{\sqrt d},\ldots,\frac{14}{\sqrt d})$   & $\mu_{3i}=(-1)^i \mu_{2i}$	\\   \hline

    {\bf Case 7  \rule{0pt}{5ex} }  &  $\mathbf{R}_1^T \mathbf{R}_1$   &  $\mathbf{R}_2^T \mathbf{R}_2$  &  $\mathbf{R}_3^T \mathbf{R}_3$	 &	$(0,0,\ldots,0)$ &  $(0,0,\ldots,0,0)$   &   $(0,0,\ldots,0,0)$\\ \hline 

    {\bf Case 8   \rule{0pt}{5ex}}  &  $\mathbf{R}_1^T \mathbf{R}_1$   &  $\mathbf{R}_2^T \mathbf{R}_2$  &  $\mathbf{R}_3^T \mathbf{R}_3$	 &	$\mathcal{N}_d(0,1)$ &  $\mathcal{N}_d(0,1)$   &   $\mathcal{N}_d(0,1)$    \\ \hline 

    {\bf Case 9   \rule{0pt}{5ex}}  &  $\mathbf{R}_1^T \mathbf{R}_1  \mathbf{R}_1^T \mathbf{R}_1$   &  $\mathbf{R}_2^T \mathbf{R}_2  \mathbf{R}_2^T \mathbf{R}_2$ &  $\mathbf{R}_3^T \mathbf{R}_3  \mathbf{R}_3^T \mathbf{R}_3$	 &	$(0,0,\ldots,0)$ &  $(0,0,\ldots,0,0)$   &   $(0,0,\ldots,0,0)$\\ \hline 

    {\bf Case 10  \rule{0pt}{5ex}}  &  $\mathbf{R}_1^T \mathbf{R}_1  \mathbf{R}_1^T \mathbf{R}_1$   &  $\mathbf{R}_2^T \mathbf{R}_2  \mathbf{R}_2^T \mathbf{R}_2$ &  $\mathbf{R}_3^T \mathbf{R}_3  \mathbf{R}_3^T \mathbf{R}_3$ &  $\mathcal{N}_d(0,1)$ &  $\mathcal{N}_d(0,1)$   &   $\mathcal{N}_d(0,1)$    \\ \hline 
\end{tabular}\vspace*{4mm}
\caption{\small Description of synthetic datasets. The above class-specific means and covariance matrices were used to sample from a multivariate Gaussian distribution. These are the same data characteristics as those  used in \cite{srivastava2007bayesian}, reflecting varying degrees of correlation between variables.  Note that there is no model mismatch between these data-sets and what is assumed by our generative models. }
\label{tab:syn}
\end{sidewaystable}

\begin{table}
\centering
    \begin{tabular}{| c  || c  | c | c | c || c | c  |  }
 \hline
 \emph{Error rate (\%)}  & $d$  &  {\em BDA7}  & {\em QB}   & {\em EDDA} & {\em model A}  & {\em model B}   \\ \hline  \hline
    {\bf Case 1\rule{0pt}{3ex}}    & 10  &  13.2   &  19.2	 &	11.2 &  12.0 $\pm$ 3.2	& 11.0 $\pm$ 2.8  \\ \hline
                  & 50  &  27.9  &  33.3	 &	21.7  &  19.9 $\pm$ 4.6  	& 15.6 $\pm$  3.4  \\ \hline
                  & 100 &  35.8  &  31.1	 &	24.8  &  32.6 $\pm$  6.0	& 19.9 $\pm$  4.3 \\ \hline \hline 

    {\bf Case 2\rule{0pt}{3ex}}    & 10  &  21.3  &  27.4	 &	16.1  &  11.9 $\pm$  3.4	& 11.4 $\pm$ 3.6 \\ \hline
                  & 50  &  26.8  &  42.6	 &	12.5  &  9.3 $\pm$  3.2	& 5.8 $\pm$ 2.2 \\ \hline
                  & 100 &  20.8  &  41.9	 &	9.0  &  26.5 $\pm$ 5.6	& 3.6 $\pm$ 2.1 \\ \hline \hline

    {\bf Case 3 \rule{0pt}{3ex}}    & 10  &  10.4  &  35.0	 &	9.1  &  27.2 $\pm$ 4.9	& 27.2 $\pm$ 5.5 \\ \hline
                  & 50  &  27.2  &  55.7	 &	21.2  &  48.6 $\pm$  5.0& 49.2 $\pm$  5.2 \\ \hline
                  & 100 &  46.9  & 56.4	 &	27.7  &  55.4	 $\pm$ 5.2 & 55.1 $\pm$ 4.9	 \\ \hline \hline 

    {\bf Case 4 \rule{0pt}{3ex}}    & 10  &  12.6  & 32.8	 &	11.6  & 11.3 $\pm$ 3.5	& 11.1 $\pm$ 4.1 \\ \hline
                  & 50  &  22.5  &  30.9	 &	17.0  &  22.5 $\pm$ 4.4	& 17.8 $\pm$ 4.0 \\ \hline
                  & 100 &  37.6  &  32.1	 &	21.1  &  30.8 $\pm$ 5.2	& 21.9 $\pm$ 4.3 \\ \hline \hline

    {\bf Case 5 \rule{0pt}{3ex}}    & 10  &  4.1  &  15.0	 &	4.4  &  12.8  $\pm$ 4.1	& 12.8 $\pm$ 3.5 \\ \hline
                  & 50  &  1.2  &  30.6	 &	0.0  &  9.2 $\pm$ 3.4	& 5.6  $\pm$ 2.7\\ \hline
                  & 100 &  0.2  &  38.3	 &	0.1  & 10.9 $\pm$ 3.8	& 5.4 $\pm$ 3.4 \\ \hline \hline 

    {\bf Case 6  \rule{0pt}{3ex}}    & 10  &  5.2  &  7.9	 &	1.7  &  4.6  $\pm$ 2.3	& 4.4  $\pm$ 2.3 \\ \hline
                  & 50  &  0.5  &  26.5	 &	0.0  &  3.9  $\pm$ 2.3	& 3.5  $\pm$ 2.4 \\ \hline
                  & 100 &  0.1  &  29.4	 &	0.0  &  4.8  $\pm$ 2.5	& 4.5  $\pm$ 2.6 \\ \hline \hline

    {\bf Case 7  \rule{0pt}{3ex}}    & 10  &  19.5  &  22.8	 &	19.7  &  20.0  $\pm$ 6.0	& 27.3  $\pm$ 7.4 \\ \hline
                  & 50  &  34.7  &  30.9	 &	63.9  &  30.2  $\pm$ 5.0 & 44.7  $\pm$ 7.8 \\ \hline
                  & 100 &  40.0  &  35.2	 &	64.8  &  35.2  $\pm$ 5.1	& 51.7  $\pm$ 7.8 \\ \hline \hline

    {\bf Case 8  \rule{0pt}{3ex}}    & 10  &  3.7  &  2.7	 &	5.1  &  1.6  $\pm$ 1.9	& 1.5  $\pm$ 1.5 \\ \hline
                  & 50  &  9.2  &  3.5	 &	25.5  &  4.4  $\pm$ 3.2	& 9.5  $\pm$ 5.0 \\ \hline
                  & 100 &  17.3  &  8.1	 &	55.2  & 8.7  $\pm$ 4.4	& 23.9  $\pm$ 9.0 \\ \hline \hline

    {\bf Case 9  \rule{0pt}{3ex}}    & 10  &  1.5  &  0.9	 &	1.0  &  0.9 $\pm$1.1	 & 5.4  $\pm$ 6.8  \\ \hline
                  & 50  &  1.3  &  0.9	 &	32.5  &  1.3 $\pm$ 1.2	& 16.9 $\pm$ 14.6 \\ \hline
                  & 100 &  2.9  &  2.8	 &	67.0  &  1.5 $\pm$ 1.5	& 22.4 $\pm$ 15.3 \\ \hline \hline

    {\bf Case 10  \rule{0pt}{3ex}}    & 10  &  0.4  &  0.1	 &	3.4  &  0.1 $\pm$ 0.6	& 0.2  $\pm$ 0.6 \\ \hline
                  & 50  &  1.7  &  0.9	 &	32.4  &  0.8 $\pm$ 1.0 	& 15.9 $\pm$ 13.6 \\ \hline
                  & 100 &  2.2  &  2.4	 &	64.0  &  1.4 $\pm$ 1.2 	& 23.4 $\pm$ 16.0 \\ \hline 

 \end{tabular}\vspace*{2mm}
\caption{\small Classification performance for synthetic datasets. Three generative Bayesian models, BDA7, QB and EDDA (results taken from \cite{srivastava2007bayesian}) are used as comparison with our models A and B. Error rates are the percentages of misclassified samples from the test data-set. The error bars for models A and B represent one standard deviation in the error rates, calculated over the 100 data realisations.}
\label{tab:syn_res}
\end{table}

The study \cite{srivastava2007bayesian} used a set of ten synthetic data-sets, all with Gaussian multivariate covariate distributions and a range of choices for class-specific means and covariance matrices. In the present study we generated data with exactly the same statistical features. The first six  of these choices were also used in \cite{friedman1989regularized}, and involve diagonal covariance matrices. The remaining four represent correlated data. Each data-set has $C=3$ outcome classes, and is separated into a training set, with $n_z=13$ samples in each class, and a validation set, with $n_z=33$ samples in each class. In terms of the balance $n_z/d$, this allows for a direct comparison with the dimensions used in \cite{srivastava2007bayesian}. The results shown in Table \ref{tab:syn_res} are all averages over 100 synthetic data runs.   The data dimensions are chosen  from $d\in\{10, 50, 100\}$.  Since all these synthetic datasets involve  multivariate Gaussian covariate distributions, there is no model mismatch with any of the models being compared.

The means and covariance matrices of the synthetic data-sets are given in Table \ref{tab:syn}. 
The covariance matrices for the correlated data-sets are defined in terms of auxiliary random $d \times d$ matrices $\mathbf{R}_z$, with i.i.d. entries sampled from the uniform distribution on the interval $[0,1]$, according to  either $\sig_z=\mathbf{R}_z^T \mathbf{R}_z$ or $\sig_z=\mathbf{R}_z^T \mathbf{R}_z  \mathbf{R}_z^T \mathbf{R}_z$. 
These covariance matrices have a single dominant eigenvalue, and  further non-dominant eigenvalues that are closer to zero for data-sets  9-10.  Data-sets 7 and 9 have all class means at the origin, whereas  each element of the class mean vectors from data-sets 8 and 10 are independently sampled from a standard normal distribution.

Table  \ref{tab:syn_res} shows the classification error rates, as  percentages of misclassified samples over the validation set.  The variability of these for results for the models BDA7, QB and EDDA, i.e. the error bars in the classification scores, is not reported in \cite{srivastava2007bayesian} (where only  the best classifier was determined using a signed ranked test). For completeness, we have included in this study the standard deviation of the error rate over the 100 synthetic data runs for our models A and B. Given that all experiments involved the same dimensions of data-sets and similar average error rates, the error bars for the \cite{srivastava2007bayesian} results are expected to be similar to those of models A and B.
We conclude from Table  \ref{tab:syn_res} that our models A and B perform on average quite similarly to the benchmark classifiers BDA7, QB and EDDA. On some data-sets model A and/or B outperform the benchmarks, on others they are outperformed. However, models A and B achieve this competitive level of classification accuracy without cross-validation, i.e. at a much lower computational cost.

\subsection{Real data}

Next we test the classification accuracy of our models against the real data-sets used in \cite{srivastava2007bayesian}, which are publicly available from the UCI machine learning repository\footnote{\url{http://archive.ics.uci.edu/ml/index.php}}.
Three data-sets were left out due to problems with matching the formats: \emph{Image segmentation} (different number of samples than \cite{srivastava2007bayesian}), \emph{Cover type} (different format of training/validation/test), and \emph{Pen digits} (different format of training/validation/test). 
Before classification, we looked for identifying characteristics which could allow for retrospective interpretation of the results, e.g. occurrence of discrete covariate values,  covariance matrix entropies, or class imbalances. None were found to be informative.

We duplicated exactly the protocol of \cite{srivastava2007bayesian}, whereby only a randomly chosen  5\% or 10\% of the samples from each class of each data-set are used for training, leaving the bulk of the data (95\% or 10\%) to serve as validation (or test) set. The resulting small training sample sizes lead to $n_z\ll  d$ for a number of data-sets, providing a rigorous test for classifiers in overfitting-prone conditions.
 For example, the set {\em Ionosphere}, with $d=34$, has original class sizes of 225 and 126 samples leading in the 5\% training scenario to training sets with $n_1=12$ and $n_2=7$. We have used the convention of rounding up any non-integer number of training samples (rounding down the number of samples  had only a minimal effect on most error rates).
The {\em baseline} column gives the classification error that would be obtained if the majority class is predicted every time. 

\begin{table}
\centering
    \begin{tabular}{| l  || l |l | l |l | l | l | l || l | l |  }
 \hline
	 \emph{Error rate (\%)} & $n$ & {\em class sizes} & $d$ & {\em baseline}  & {\em BDA7}  & {\em QB}   & {\em EDDA} & {\em model A}  & {\em model B} \\ \hline  \hline
    {\bf Heart}    &  270 & 150,120 & 10 & 44.4 & 27.4  &  32.0	 &	28.3  &  30.3	& 30.1 \\ \hline \hline

    {\bf Ionosphere}  &  351 & 225, 126 & 34  & 35.9&  12.5  &  11.1	 &	23.3  &  8.3	& 7.5 \\  \hline \hline

    {\bf Iris}    & 150 & 50,50,50 & 4 & 66.6 &  6.2  &  5.9	 &	7.4  & 7.5	& 6.6 \\ \hline \hline

    {\bf Pima}    &  768 & 500,268 &  8 & 34.9 &  28.4  &  29.7	 &	29.0  &  28.8	& 28.9 \\ \hline \hline

    {\bf Sonar}    &  208 & 97,111 & 60 & 46.6 & 31.2  &  33.7	 &	34.8  &  34.9	& 33.8 \\  \hline \hline

    {\bf Thyroid}   &  215 & 150,35,30 & 5  & 30.2 &  7.9  &  9.1	 &	8.6  &  7.6	& 7.9 \\  \hline \hline

    {\bf Wine}   &  178& 59,71,48 &13 & 60.1 & 7.9  &  16.9	 &	8.2  &  15.6  & 16.0 \\  \hline 

 \end{tabular}\vspace*{2mm}
\caption{\small Average error rate using randomly selected 10\% of training samples in each class. The remaining 90\% of samples were used as a validation set. Error rates are the percentage of misclassified samples over this validation set.}
\label{tab:ten_percent}
\end{table}

\begin{table}
\centering
    \begin{tabular}{| l  || l | l |l | l | l | l | l || l | l |  }
 \hline
 \emph{Error rate (\%)} & $n$ &{\em class sizes}  & $d$ &   {\em baseline} & {\em BDA7}  & {\em QB}   & {\em EDDA} & {\em model A}  & {\em model B} \\ \hline  \hline
    {\bf Heart}    & 270 & 150,120 & 10  & 44.4 &  30.6  &  38.5	 &	33.9  &  38.8 	&  39.6  \\ \hline \hline

    {\bf Ionosphere}  & 351 & 225, 126 & 34 & 35.9&  16.9  &  16.1	 &	26.0  &  10.3	& 8.8  \\  \hline \hline

    {\bf Iris}    & 150 & 50,50,50 & 4& 66.6 & 6.9  &  7.6	 &	9.40  &  12.8&  11.4 \\ \hline \hline

    {\bf Pima}    &   768 & 500,268 & 8 & 34.9&  29.7  & 32.7 &	30.7  & 30.3	& 30.8  \\ \hline \hline

    {\bf Sonar}    & 208 & 97,111 & 60 & 46.6 & 36.8  &  40.4	 &	39.8  & 45.6 & 39.0  \\  \hline \hline

    {\bf Thyroid}   &  215 & 150,35,30 & 5& 30.2 & 11.7  &  14.8	 &	14.7  &   34.5 	& 14.6 \\  \hline \hline

    {\bf Wine}   &   178 & 59,71,48 & 13 & 60.1& 9.6  &  33.1	 &	11.2  &  54.4 &  33.0 \\  \hline 

 \end{tabular}\vspace*{2mm}
\caption{\small  Average error rate using randomly selected 5\% of training samples in each class. The remaining 95\% of samples were used as a validation set. Error rates are the percentage of misclassified samples over this validation set.}
\label{tab:five_percent}
\end{table}

We conclude from the classification results shown in Tables \ref{tab:ten_percent} and \ref{tab:five_percent} (which are to be interpreted as having non-negligible error bars), that also 
for the real data, models A and B are competitive with the other Bayesian classifiers. 
The exceptions are {\em Ionosphere} (where models A and B outperform the benchmark methods, in both tables) and the data-sets {\em Thyroid} and {\em Wine} (where in Table \ref{tab:five_percent}  our model A is being outperformed). Note that in Table \ref{tab:five_percent}, \emph{Thyroid} and \emph{Wine} have only 2 or 3 data samples in some classes of the training set. This results in nearly degenerate class-specific covariance matrices,  which hampers the optimization of hyperparameters via evidence maximization. Model B behaves well even in these tricky cases, presumably due to the impact of its additional hyperparameter $\gamma_{0z}=d/\hat{\bX}^2_z$.
As expected, upon testing classification performance using leave-one-out cross-validation (details not shown here) rather than the 5\% or 10\% training set methods above, all error rates are significantly lower.

\section{Discussion}
\label{sec:discussion}

 In this paper we considered generative models for supervised Bayesian classification in high-dimensional spaces. Our aim was to derive expressions for the optimal hyperparameters and predictive probabilities in closed form. Since the dominant cause of overfitting in the classification of high-dimensional data is using point estimates for high-dimensional parameter vectors, we believe that by carefully choosing Bayesian models for which parameter integrals  are analytically tractable, we will need point estimates only at  hyperparameter level, reducing overfitting.

We showed that the standard priors  of  Bayesian classifiers that are based on class-specific multivariate Gaussian covariate distributions can be  generalized, from which we derive two special model cases (A and B) for which predictive probabilities can be derived analytically in fully explicit form. Model A is known in the literature as Quadratic Bayes \cite{brown1999discrimination}, whereas model B is novel and has not yet appeared in the literature.  
In contrast to common practice for most Bayesian classifiers, we  use evidence maximization \cite{mackay1999comparison} to find analytical expressions for our hyperparameters in both models. This allows us to find their optimal values without needing to resort to computationally expensive cross-validation protocols.

We found that the alternative (but significantly faster) hyperparameter determination by evidence maximization leads to hyperparameters that are generally very similar to those obtained via cross-validation, and that the classification performance of our models A and B degrades only gracefully in the `dangerous' regime $n\ll d$ where we would expect extreme overfitting. 
We compared the classification performance of our models on the extensive synthetic and real data-sets that have been used earlier as performance benchmarks  in \cite{srivastava2006distribution,srivastava2007bayesian}. Interestingly, the performance of our models A and B turned out to be competitive with state-of-the-art Bayesian models that use cross-validation, despite the large reduction in computational expense. This will enable users in practice to classify high-dimensional data-sets quicker, without compromising on accuracy.
\\[5mm]
{\bf Acknowledgements}
\\[2mm]
This work was supported by the Biotechnology and Biological Sciences Research Council (UK) and by GlaxoSmithKline Research and Development Ltd under grant BIDS3000023807.

\end{document}